\documentclass[aps,pra,twocolumn,eqsecnum,superscriptaddress]{revtex4-1}
\usepackage{graphicx}% Include figure files
\usepackage{color}
\usepackage{amsmath}
\usepackage{amssymb}
\usepackage{amsthm}
\usepackage{hyperref}
\usepackage{enumitem}
\usepackage{bm}

\newcommand{\rmd}{\text{d}}
\newcommand{\rme}{\text{e}}
\newcommand{\rmi}{\text{i}}

\renewcommand{\Re}{\mathop{\text{Re}}\nolimits}
\renewcommand{\Im}{\mathop{\text{Im}}\nolimits}

\definecolor{dgreen}{rgb}{0,0.5,0}

\definecolor{delete}{cmyk}{0.5,0,0,0}

\begin{document}
%\preprint{}

\title{Relaxation to Gaussian Generalized Gibbs Ensembles in Quadratic Bosonic Systems in the Thermodynamic Limit}


%\author{}
%\thanks{}
%\affiliation{}
\author{Takaaki Monnai}
\affiliation{Department of Materials and Life Science, Seikei University, Tokyo 180-8633, Japan}
\author{Shohei Morodome}
\affiliation{Department of Physics, Waseda University, Tokyo 169-8555, Japan}
\author{Kazuya Yuasa}
\affiliation{Department of Physics, Waseda University, Tokyo 169-8555, Japan}


%\date{\today}
\date[]{August 5, 2019}

\begin{abstract}
Integrable quantum many-body systems are considered to equilibrate to generalized Gibbs ensembles (GGEs) characterized by the expectation values of integrals of motion.
We study the dynamics of exactly solvable quadratic bosonic systems in the thermodynamic limit, and show a general mechanism for the relaxation to GGEs, in terms of the diagonal singularity.
We show analytically and explicitly that a free bosonic system relaxes from a general (not necessarily Gaussian) initial state under certain physical conditions to a Gaussian GGE\@.
We also show the relaxation to a Gaussian GGE in an exactly solvable coupled system, a harmonic oscillator linearly interacting with bosonic reservoirs.
\end{abstract}

%\keywords{}

\maketitle

\section{Introduction}
\label{sec:Introduction} 
Recent advances in experimental studies of relaxation processes with quantum atomic gases \cite{Kinoshita1,Kinoshita2,Gring1,Schmiedmayer1,Langen1,Kaufman1} have been stimulating theoretical studies on equilibration and prethermalization of isolated quantum many-body systems \cite{ref:PolkovnikovRMP,Yukalov1,ref:EisertFriesdorfGogolin-ReviewNatPhys,ref:GooldReviewJPA,Goglin1,Vidmar1,ref:CalabresePhysicaA,ref:MoriIkedaKaminishiUeda-ReviewJPB}. 
Among various issues, it is recognized that an integrable large quantum system would relax to a nonthermal steady state described by a \textit{generalized Gibbs ensemble (GGE)} \cite{Rigol2,ref:CramersDawsonEisertOsbornePRL2008,Rigol1,Caux1,Wouters1}, which is characterized by the expectation values of a set of integrals of motion.

In Ref.\ \cite{Rigol2}, the GGE was conjectured on the basis of the principle of maximum entropy \cite{Jaynes1} and verified numerically, and then, in Refs.\ \cite{Caux1,Wouters1}, it was studied with integrable systems under initial quench scenarios.
There exist many studies on this issue: see reviews \cite{ref:PolkovnikovRMP,ref:EisertFriesdorfGogolin-ReviewNatPhys,ref:GooldReviewJPA,Goglin1,Vidmar1,ref:CalabresePhysicaA,ref:MoriIkedaKaminishiUeda-ReviewJPB} and references therein.
Nonetheless, it remains a challenging open problem to clarify a concrete scenario of the microscopic mechanism of the relaxation to GGE, e.g.\ in the context of the Liouville integrability in Hamiltonian systems, where we essentially perform canonical transformations to an assembly of noninteracting oscillators, i.e.\ action-angle variables \cite{ref:KuboTextbookI-Chap5}.
For the unitary evolution of nonintegrable systems, on the other hand, there are intensive studies, e.g.\ to give foundations to the quantum ergodic theorem \cite{vonNeumann1,vonNeumann2,Typicality1,Goldstein2,Typicality3,Typicality2} and to understand the relaxation times \cite{relaxation1,Goldstein3,Monnai2,relaxation2,Goldstein1b,relaxation3}.

In Ref.\ \cite{ref:CramerEisert-CentralLimitTheorem} (see also Ref.\ \cite{ref:CramersDawsonEisertOsbornePRL2008}), it was shown that a one-dimensional bosonic lattice system with a quadratic Hamiltonian with nearest-neighbor hopping terms locally equilibrates from an arbitrary initial state fulfilling conditions such as clustering and the absence of anomalous correlations to a \textit{Gaussian} GGE for a finite time duration. 
Similarly, in Ref.\ \cite{ref:GogolinEisert-GaussificationFermi}, the local equilibration to a Gaussian GGE during a finite time interval was shown for a fermionic $d_\mathcal{L}$-dimensional cubic lattice system with a quadratic Hamiltonian with finite-range interactions, under a clustering condition on the initial state. 
The key to the proofs is the Lieb-Robinson bound \cite{Goglin1}.
In addition, power-law relaxation towards Gaussian GGE was studied for similar bosonic/fermionic lattice models with quadratic Hamiltonians, by making use of the Kusmin-Landau bound \cite{Farrelly1}  and  the stationary-phase approximation \cite{Murthy1} under the clustering condition on the initial state. 
See also Ref.\ \cite{Sotiriadis1} for further arguments on the clustering decomposition in the context of Gaussian GGE\@.
These results are interesting because it is shown that the systems equilibrate to the GGEs \textit{during time intervals} (not simply \textit{on average in time} \cite{Goglin1}) and that only subsets of the sets of conserved quantities of the integrable systems are relevant to the GGEs.

In this paper, we provide another contribution to this issue, studying \textit{exactly solvable quadratic bosonic models}, namely, a class of models which can be mapped to free bosonic fields in the thermodynamic limit. 
The existence of such a mapping corresponds to the Liouville integrability for classical systems \cite{note:QuantumIntegrability}. %\cite{ref:Caux-QuantumIntegrability,ref:Braak-RabiExact}.
We solve the evolution of the state of the whole system in the thermodynamic limit exactly and explicitly, and observe that it relaxes to a simple Gaussian GGE in the long-time limit, under certain conditions on the initial state.

More specifically, we are going to show the following.
We consider solvable bosonic systems, with canonical normal modes say $\hat{b}_{\bm{k}}$, in the thermodynamic limit in $D$-dimensional space.
We assume the following physical conditions on the initial state.
\begin{enumerate}%[leftmargin=*]
\item[(i)]
The initial state of the system, which is non-Gaussian in general, is prepared irrespective of the Hamiltonian of the system (like in a quench scenario).
\item[(ii)]
We allow the correlations in the initial state to possess translationally invariant components (particles can be distributed all over the space and correlations can exist everywhere in space in the initial state).
\item[(iii)]
But the correlations in the initial state are assumed to be of finite range.
\item[(iv)]
The first moment $\langle\hat{b}_{\bm{k}}\rangle$ in the initial state is assumed to be nonvanishing only locally, i.e.\ free from a translationally invariant component.
\end{enumerate}
In addition:
\begin{enumerate}%[leftmargin=*]
\item[(v)]
We exclude observables extending all over the space, since the expectation values of such quantities diverge in the thermodynamic limit and are not actually measurable.
\end{enumerate}
Under these conditions, we show that the system relaxes to a Gaussian GGE of the form
\begin{equation}
\hat{\rho}_\text{GGE}
\propto
\exp\!\left(
-\int \rmd^D\bm{k}
\ln(1+f_{\bm{k}}^{-1})
\hat{b}_{\bm{k}}^\dag
\hat{b}_{\bm{k}}
\right)
\label{eqn:GGE}
\end{equation}
in the long-time limit, where $f_{\bm{k}}$ is the translationally invariant component of the single-particle correlation $\langle (\hat{b}_{\bm{k}}^\dag-\langle\hat{b}_{\bm{k}}^\dag\rangle)(\hat{b}_{\bm{k}'}-\langle\hat{b}_{\bm{k}'}\rangle)\rangle$ in the initial state. 
Observe that the GGE in (\ref{eqn:GGE}) admits only the occupation numbers $\hat{I}_{\bm{k}}=\hat{b}_{\bm k}^\dagger\hat{b}_{\bm k}$ as integrals of motion (cf.\ \cite{Sels1,Polkovnikov1}).
The state $\hat{\rho}_\text{GGE}$ in (\ref{eqn:GGE}) is a \textit{Gaussian state}, since its Wigner function is Gaussian \cite{ref:ContVarQI,ref:GaussianQI}.
Therefore, we call the state $\hat{\rho}_\text{GGE}$ in (\ref{eqn:GGE}) \textit{Gaussian GGE}\@.
The correlations among the occupation numbers such as $I_{\bm{k}}I_{\bm{k}'}$ do not survive in the equilibrium state $\hat{\rho}_\text{GGE}$.

In contrast to the previous works \cite{ref:CramerEisert-CentralLimitTheorem} and \cite{ref:GogolinEisert-GaussificationFermi}, in which the results are rigorously proved for large but finite lattice systems, we directly go to the thermodynamic limit and deal with field-theoretical Hamiltonians with continuous spectrum.
This greatly simplifies the analysis and the picture.
Our results are valid for any spatial dimension, anomalous correlations are allowed in the initial states, and we do not focus on a part of the system (we do not take partial trace) \cite{Note1}.
In this approach, the relaxation to a GGE is understood in terms of the \textit{diagonal singularity} \cite{VANHOVE1955901,ANTONIOU1997737}, or equivalently, as a consequence of the Riemann-Lebesgue lemma \cite{ref:FourierAnalysisKoerner}, smearing the spectrum due to the rapid oscillations in the long-time limit.
We will see that some ``regularity'' of the correlation functions and the observables is important, for the Riemann-Lebesgue lemma to work.
In particular, we stress that observables are relevant to the relaxation process (e.g.\ for the relaxation time).

We first study the dynamics of a free bosonic field in Sec.\ \ref{sec:FreeBoson}\@.
We show the relaxation from Gaussian initial states to the Gaussian GGE in (\ref{eqn:GGE}), and generalize the result to non-Gaussian initial states.
We also study a solvable coupled system, i.e.\ a harmonic oscillator interacting with bosonic reservoirs, in Sec.\ \ref{sec:Friedrichs}\@.
We show the relaxation from factorized initial states to a Gaussian GGE like (\ref{eqn:GGE}), and generalize the result to correlated initial states.
Concluding remarks are given in Sec.\ \ref{sec:Conclusions}, and some involved calculations demonstrating the decay of the cumulants leading to the Gaussification and the proofs of some mixing properties of the Gaussian GGE, which are key to the relaxation, are shown in Appendices \ref{app:DecayCumulants}--\ref{app:RelaxCorrelated}\@.

The equilibration of an exactly solvable quadratic model similar to the one we study in Sec.\ \ref{sec:Friedrichs} was discussed in Ref.\ \cite{ref:EquilibrationExact-PagelAlvermannFehskePRE} on the basis of its exact solution.
But only the reduced dynamics of the central harmonic oscillator, with the reservoirs' degrees of freedom traced out, was analyzed there with a factorized initial state with the reservoirs being in a Gaussian state.
In Sec.\ \ref{sec:Friedrichs}, in contrast, we will analyze the evolution of the state of the whole system, including the reservoirs, from a correlated initial state, which is not Gaussian in general.
The total system relaxes to a GGE like (\ref{eqn:GGE}).

\section{Free Bosonic Field}
\label{sec:FreeBoson}
Let us start with a free bosonic field in $D$-dimensional space.
The annihilation and creation operators $\hat{b}_{\bm{k}}$ and $\hat{b}_{\bm{k}}^\dag$ for bosons with momentum $\bm{k}$ obey the canonical commutation relations
$[\hat{b}_{\bm{k}},\hat{b}_{\bm{k}'}]=0$, 
$[\hat{b}_{\bm{k}},\hat{b}_{\bm{k}'}^\dag]=\delta^D(\bm{k}-\bm{k}')$,
and the Hamiltonian of the system is given by
\begin{equation}
\hat{H}=\int \rmd^D\bm{k}\,\omega_k\hat{b}_{\bm{k}}^\dag\hat{b}_{\bm{k}},
\label{eqn:FreeBoson}
\end{equation}
with a dispersion relation $\omega_k\ge0$.
We set $\hbar=1$.

\subsection{Gaussian Initial State}
\label{sec:FreeBosonGaussian}
We first focus on Gaussian initial states \cite{ref:ContVarQI,ref:GaussianQI}.
The density operator for a Gaussian state is formally given by \cite{ref:GaussFidelity}
\begin{widetext}
\begin{equation}
\hat{\rho}_0
\propto
\exp\!\left[
-\frac{1}{2}
\int\rmd^D\bm{k}\int\rmd^D\bm{k}'
\begin{pmatrix}
\hat{b}_{\bm{k}}^\dag-\langle\hat{b}_{\bm{k}}^\dag\rangle
&
\hat{b}_{\bm{k}}-\langle\hat{b}_{\bm{k}}\rangle
\end{pmatrix}
\Theta_{\bm{k}\bm{k}'}
\begin{pmatrix}
\medskip
\hat{b}_{\bm{k}'}-\langle\hat{b}_{\bm{k}'}\rangle
\\
\hat{b}_{\bm{k}'}^\dag-\langle\hat{b}_{\bm{k}'}^\dag\rangle
\end{pmatrix}
\right],
\label{eqn:DensityOp}
\end{equation}
\end{widetext}
where $\langle{}\cdots{}\rangle$ denotes the expectation value in the state $\hat{\rho}_0$, and $\Theta_{\bm{k}\bm{k}'}$ is a $2\times2$ matrix satisfying $(\Theta_{\bm{k}\bm{k}'})^\dag=\Theta_{\bm{k}'\bm{k}}$.
Note however that, for an infinitely extended system like the present bosonic field in the thermodynamic limit, such a density operator is not normalizable and is mathematically ill-defined. 
Instead, in the $C^*$-algebraic approach to quantum field theory and quantum statistical mechanics \cite{Bratteli1,ref:NESS-AschbacherPillet-JSP,ref:NESS-AschbacherJaksicPautratPillet,ref:NESS-Tasaki}, states are rigorously characterized by characteristic functionals, i.e.\ the generating functionals of correlation functions.
The characteristic functional of a Gaussian state is Gaussian, and for the present bosonic field it reads as
\begin{widetext}
\begin{align}
\chi_0[J,J^*]
&=
\langle
\rme^{\int\rmd^D\bm{k}\,(J_{\bm{k}}\hat{b}_{\bm{k}}^\dag-J_{\bm{k}}^*\hat{b}_{\bm{k}})}
\rangle
\nonumber
\displaybreak[0]
\\
&=
\exp\!\left[
-\frac{1}{2}
\int\rmd^D\bm{k}\int\rmd^D\bm{k}'
\begin{pmatrix}
J_{\bm{k}}^*
&
J_{\bm{k}}
\end{pmatrix}
V_{\bm{k}\bm{k}'}
\begin{pmatrix}
\medskip
J_{\bm{k}'}
\\
J_{\bm{k}'}^*
\end{pmatrix}
+\int\rmd^D\bm{k}\,(J_{\bm{k}}\langle\hat{b}_{\bm{k}}^\dag\rangle-J_{\bm{k}}^*\langle\hat{b}_{\bm{k}}\rangle)
\right],
\label{eqn:GaussChar}
\end{align}
where $V_{\bm{k}\bm{k}'}$ is the covariance matrix of the Gaussian state, defined by
\begin{equation}
V_{\bm{k}\bm{k}'}
=
\begin{pmatrix}
\medskip
\frac{1}{2}
\langle
\{\hat{b}_{\bm{k}}-\langle\hat{b}_{\bm{k}}\rangle,\hat{b}_{\bm{k}'}^\dag-\langle\hat{b}_{\bm{k}'}^\dag\rangle\}\rangle
&
-\langle
(\hat{b}_{\bm{k}}-\langle\hat{b}_{\bm{k}}\rangle)(\hat{b}_{\bm{k}'}-\langle\hat{b}_{\bm{k}'}\rangle)\rangle
\\
-\langle
(\hat{b}_{\bm{k}}^\dag-\langle\hat{b}_{\bm{k}}^\dag\rangle)(\hat{b}_{\bm{k}'}^\dag-\langle\hat{b}_{\bm{k}'}^\dag\rangle)\rangle
&
\frac{1}{2}
\langle
\{\hat{b}_{\bm{k}}^\dag-\langle\hat{b}_{\bm{k}}^\dag\rangle,\hat{b}_{\bm{k}'}-\langle\hat{b}_{\bm{k}'}\rangle\}\rangle
\end{pmatrix}
\label{eqn:CovarianceMatrix}
\end{equation}
\end{widetext}
and satisfying $(V_{\bm{k}\bm{k}'})^\dag=V_{\bm{k}'\bm{k}}$.
The covariance matrix $V_{\bm{k}\bm{k}'}$ is formally related to the matrix $\Theta_{\bm{k}\bm{k}'}$ in the Gaussian density operator (\ref{eqn:DensityOp}) through 
\begin{equation}
\Theta
=2Z\coth^{-1}(2ZV)
=Z\ln\frac{1+(2ZV)^{-1}}{1-(2ZV)^{-1}}
\label{eqn:GaussianMatrix1}
\end{equation}
with
$
Z
=\begin{pmatrix}
1&0\\
0&-1
\end{pmatrix}$
\cite{ref:GaussFidelity}, where $\Theta$ and $V$ are understood as infinite-dimensional matrices with their rows and columns labeled by the continuous indices $\bm{k}$ and $\bm{k}'$ as well \cite{note:Matrix}.

We impose physical conditions on the initial state $\hat{\rho}_0$.
We assume that the correlation functions involved in the covariance matrix $V_{\bm{k}\bm{k}'}$ are endowed with the following structures:
\begin{align}
\langle
(\hat{b}_{\bm{k}}^\dag-\langle\hat{b}_{\bm{k}}^\dag\rangle)
(\hat{b}_{\bm{k}'}-\langle\hat{b}_{\bm{k}'}\rangle)
\rangle
&=
f_{\bm{k}}\delta^D(\bm{k}-\bm{k}')
+F_{\bm{k}\bm{k}'},
\label{eqn:F}
\displaybreak[0]\\
\langle
(\hat{b}_{\bm{k}}-\langle\hat{b}_{\bm{k}}\rangle)
(\hat{b}_{\bm{k}'}-\langle\hat{b}_{\bm{k}'}\rangle)
\rangle
&=
g_{\bm{k}}\delta^D(\bm{k}+\bm{k}')
+G_{\bm{k}\bm{k}'},
\label{eqn:G}
\end{align}
where $f_{\bm{k}}$, $F_{\bm{k}\bm{k}'}$, $g_{\bm{k}}$, and $G_{\bm{k}\bm{k}'}$ are ``regular'' functions of $\bm{k}$ and $\bm{k}'$ (so that the Riemann-Lebesgue lemma works later \cite{note:RiemannLebesgueLemma}). 
We also assume that the first moment $\langle\hat{b}_{\bm{k}}\rangle$ is regular in $\bm{k}$.
For instance, in the case of the canonical ensemble $\hat{\rho}_0\propto \rme^{-\hat{H}/k_BT}$, we have $f_{\bm{k}}=1/(\rme^{\omega_k/k_BT}-1)$, which is the Bose distribution function, with the other components $F_{\bm{k}\bm{k}'}$, $g_{\bm{k}}$, $G_{\bm{k}\bm{k}'}$, and $\langle\hat{b}_{\bm{k}}\rangle$ vanishing.
We are however interested in more general Gaussian states than the canonical ensemble.

The first contribution $f_{\bm{k}}\delta^D(\bm{k}-\bm{k}')$  to the correlation (\ref{eqn:F}) represents the translationally invariant component in space, in the corresponding correlation function $\langle[\hat{\psi}^\dag(\bm{r})-\langle\hat{\psi}^\dag(\bm{r})\rangle][\hat{\psi}(\bm{r}')-\langle\hat{\psi}(\bm{r}')\rangle]\rangle$ in the configuration space, where $\hat{\psi}(\bm{r})$ is the field operator defined by $\hat{\psi}(\bm{r})=\int\rmd^D\bm{k}\,\hat{b}_{\bm{k}}\rme^{-\rmi\bm{k}\cdot\bm{r}}/\sqrt{(2\pi)^D}$.
In particular, it yields a particle distribution $\langle\hat{\psi}^\dag(\bm{r})\hat{\psi}(\bm{r})\rangle$ uniform over the space.
Note that in general a sharp and narrow spectrum in the momentum space like the delta function in (\ref{eqn:F}) corresponds to a widely spread distribution in the configuration space.
The second contribution $F_{\bm{k}\bm{k}'}$, on the other hand, adds a nonuniformity to the particle distribution, and rules the single-particle coherence.
The regularity of $F_{\bm{k}\bm{k}'}$ implies that the coherence length of the single-particle correlation is finite, i.e.\ $\langle[\hat{\psi}^\dag(\bm{r})-\langle\hat{\psi}^\dag(\bm{r})\rangle][\hat{\psi}(\bm{r}')-\langle\hat{\psi}(\bm{r}')\rangle]\rangle$ decays as $|\bm{r}-\bm{r}'|$ increases.
In other words, particles are distributed all over the space (the total number of particles is infinite), while the single-particle coherence length is finite.

Similarly, the first contribution $g_{\bm{k}}\delta^D(\bm{k}+\bm{k}')$ to the other correlation (\ref{eqn:G}) represents the translationally invariant component in the pair correlation $\langle[\hat{\psi}(\bm{r})-\langle\hat{\psi}(\bm{r})\rangle][\hat{\psi}(\bm{r}')-\langle\hat{\psi}(\bm{r}')\rangle]\rangle$ in the configuration space.
This pair correlation decays as $|\bm{r}-\bm{r}'|$ increases, due to the regularity of $G_{\bm{k}\bm{k}'}$.
That is, the pairing is allowed everywhere in space (the total number of pairs can be infinite), while the size of each pair is finite.

Under these conditions, the covariance matrix (\ref{eqn:CovarianceMatrix}) of the initial Gaussian state is reduced to
\begin{widetext}
\begin{equation}
V_{\bm{k}\bm{k}'}
=
\begin{pmatrix}
\medskip
(\frac{1}{2}+f_{\bm{k}})\delta^D(\bm{k}-\bm{k}')
+
F_{\bm{k}\bm{k}'}^*
&
-g_{\bm{k}}\delta^D(\bm{k}+\bm{k}')
-G_{\bm{k}\bm{k}'}
\\
-g_{\bm{k}}^*\delta^D(\bm{k}+\bm{k}')
-G_{\bm{k}\bm{k}'}^*
&
(\frac{1}{2}+f_{\bm{k}})\delta^D(\bm{k}-\bm{k}')
+
F_{\bm{k}\bm{k}'}
\end{pmatrix}.
\label{eqn:CovMatStruc}
\end{equation}
\end{widetext}

Finally, the regularity of $\langle\hat{b}_{\bm{k}}\rangle$ means that a nonvanishing first moment $\langle\hat{\psi}(\bm{r})\rangle$ is allowed only locally in space.
These are our physical conditions on the initial state $\hat{\rho}_0$.
The particles are distributed all over the space, while the correlations are assumed to be of finite range, excluding peculiar long-range correlations.

We note some conditions that have to be satisfied by $f_{\bm{k}}$, $F_{\bm{k}\bm{k}'}$, $g_{\bm{k}}$, and $G_{\bm{k}\bm{k}'}$. 
First, it is clear from the definitions of these functions in (\ref{eqn:F}) and (\ref{eqn:G}) that they possess the symmetries 
$f_{\bm{k}}=f_{\bm{k}}^*$, 
$F_{\bm{k}\bm{k}'}=F_{\bm{k}'\bm{k}}^*$, 
$G_{\bm{k}\bm{k}'}=G_{\bm{k}'\bm{k}}$.
Second, the covariance matrix $V_{\bm{k}\bm{k}'}$ should satisfy $V+Z/2\ge0$ as an infinite-dimensional matrix with its rows and columns labeled by the continuous indices $\bm{k}$ and $\bm{k}'$ as well, and with $Z$ here representing $Z\delta^D(\bm{k}-\bm{k}')$. 
It is an uncertainty relation expressed in terms of the covariance matrix \cite{ref:ContVarQI,ref:GaussianQI}.

\subsection{Relaxation to GGE}
We are now ready to study the dynamics of the system evolving according to the Hamiltonian (\ref{eqn:FreeBoson}) from the Gaussian initial state (\ref{eqn:GaussChar}) with the covariance matrix (\ref{eqn:CovMatStruc}).
In the Heisenberg picture, the characteristic functional of the state at time $t$ is calculated as 
\begin{align}
\chi_t[J,J^*]
&=
\langle
\rme^{\int \rmd^D\bm{k}\,(J_{\bm{k}}\hat{b}_{\bm{k}}^\dag-J_{\bm{k}}^*\hat{b}_{\bm{k}})}
\rangle_t
\nonumber
\displaybreak[0]
\\
&=
\langle
\rme^{\int \rmd^D\bm{k}\,(J_{\bm{k}}\hat{b}_{\bm{k}}^\dag \rme^{\rmi\omega_kt}-J_{\bm{k}}^*\hat{b}_{\bm{k}}\rme^{-\rmi\omega_kt})}
\rangle
\nonumber
\displaybreak[0]
\\
&
=
\chi_0[J\rme^{\rmi\omega t},J^*\rme^{-\rmi\omega t}],
\label{Characteristic1}
\end{align}
where $\langle{}\cdots{}\rangle_t$ denotes the expectation value in the state $\hat{\rho}(t)=\rme^{-\rmi\hat{H}t}\hat{\rho}_0\rme^{\rmi\hat{H}t}$ at time $t$.
For the Gaussian initial state (\ref{eqn:GaussChar}) with the covariance matrix (\ref{eqn:CovMatStruc}), it reads
\begin{widetext}
\begin{align}
\chi_t[J,J^*]
=
\exp\biggl(
&
{-\frac{1}{2}}
\int \rmd^D\bm{k}\,[
(1+2f_{\bm{k}})|J_{\bm{k}}|^2
-2\Re(g_{\bm{k}}^*J_{\bm{k}}J_{-\bm{k}}\rme^{2\rmi\omega_kt})
]
\nonumber\displaybreak[0]\\
&
{}-
\int \rmd^D\bm{k}
\int \rmd^D\bm{k}'\,[
J_{\bm{k}}F_{\bm{k}\bm{k}'}J_{\bm{k}'}^*\rme^{\rmi(\omega_k-\omega_{k'})t}
-\Re(J_{\bm{k}}G_{\bm{k}\bm{k}'}^*J_{\bm{k}'}\rme^{\rmi(\omega_k+\omega_{k'})t})
]
+2\rmi\Im\int \rmd^D\bm{k}\,J_{\bm{k}}\rme^{\rmi\omega_kt}\langle\hat{b}_{\bm{k}}^\dag\rangle
\biggr).
\label{eqn:GaussCharEvolve}
\end{align}
\end{widetext}
In the long-time limit $t\to\infty$, the system relaxes to
\begin{align}
\chi_t[J,J^*]
&\xrightarrow{t\to\infty}
\exp\!\left(
-\frac{1}{2}
\int \rmd^D\bm{k}\,
(1+2f_{\bm{k}})|J_{\bm{k}}|^2
\right)
\nonumber
\displaybreak[0]
\\
&
=\chi_\text{GGE}[J,J^*],
\label{Characteristic2}
\end{align}
by the Riemann-Lebesgue lemma \cite{note:RiemannLebesgueLemma}.
In terms of the density operator, it formally means
\begin{equation}
\hat{\rho}(t)
\xrightarrow{t\to\infty}
\hat{\rho}_\text{GGE}
\propto
\exp\!\left(
-\int \rmd^D\bm{k}
\ln(1+f_{\bm{k}}^{-1})
\hat{b}_{\bm{k}}^\dag
\hat{b}_{\bm{k}}
\right),
\label{density1}
\end{equation}
recalling the conversion formula (\ref{eqn:GaussianMatrix1}) applied to the diagonal covariance matrix $V_{\bm{k}\bm{k}'}^{(\text{GGE})}=(\frac{1}{2}+f_{\bm{k}})\delta^D(\bm{k}-\bm{k}')\openone_2$ in (\ref{Characteristic2}).
If $f_{\bm{k}}$ is the Bose distribution function $f_{\bm{k}}=1/(\rme^{\omega_k/k_BT}-1)$, this equilibrium state coincides with the canonical ensemble $\hat{\rho}_\text{can}\propto \rme^{-\hat{H}/k_BT}$ at temperature $T$.
If $f_{\bm{k}}$ is different from the Bose distribution function, the equilibrium state $\hat{\rho}_\text{GGE}$ is a GGE, with a set of integrals of motion $\hat{I}_{\bm{k}}=\hat{b}_{\bm{k}}^\dag\hat{b}_{\bm{k}}$.
That is why we have named the equilibrium density operator $\hat{\rho}_\text{GGE}$ in (\ref{density1}).

The mechanism for the relaxation to the GGE in this simple example is clear.
It is due to the Riemann-Lebesgue lemma with the \textit{diagonal singularity} $\delta^D(\bm{k}-\bm{k}')$ \cite{VANHOVE1955901,ANTONIOU1997737} in the correlation (\ref{eqn:F}) in the covariance matrix $V_{\bm{k}\bm{k}'}$ of the initial Gaussian state.
The translationally invariant component with the delta function $\delta^D(\bm{k}-\bm{k}')$ in the normal correlation (\ref{eqn:F}) survives in the long-time limit, while the other components decay away.
We notice that the equilibration time depends not only on $F_{\bm{k}\bm{k}'}$, $g_{\bm{k}}$, $G_{\bm{k}\bm{k}'}$, and $\langle\hat{b}_{\bm{k}}\rangle$, characterizing the initial state, but also on $J_{\bm{k}}$, related to observables.
In taking the limit in (\ref{Characteristic2}), we have assumed the regularity of $J_{\bm{k}}$ so that the Riemann-Lebesgue lemma works.
It physically means that our observables are assumed to be spatially localized (of finite size) \cite{note:Locality}.
If, for instance, one considers a nonlocal observable spreading over a very large region in space, it corresponds to taking a very narrow function $J_{\bm{k}}$ in the momentum space, and the relaxation due to the Riemann-Lebesgue lemma becomes very slow.
In this way, the time scale for the relaxation to the GGE is ruled by the locality of the observables of interest ($J_{\bm{k}}$), as well as the locality of the correlations in the initial state ($F_{\bm{k}\bm{k}'}$, $g_{\bm{k}}$, $G_{\bm{k}\bm{k}'}$, and $\langle\hat{b}_{\bm{k}}\rangle$).

In Appendix \ref{app:DecayCumulants}, we provide explicit formulas for the time evolutions of the correlations for Gaussian spectra with a quadratic dispersion relation $\omega_k$.
The correlations actually decay except for the translationally invariant component of the normal correlation.

\subsection{Non-Gaussian Initial State}
\label{sec:NonGaussian}
We have so far focused on Gaussian initial states.
Let us generalize the analysis to non-Gaussian initial states.
The non-Gaussianity is characterized by the higher-order cumulants in the characteristic functional $\chi_0[J,J^*]$ of the initial state.
For instance, suppose that there exists a third-order cumulant like 
\begin{align}
&\ln\chi_0[J,J^*]
\nonumber
\displaybreak[0]
\\
&\ \ %
={}\cdots
{}+\int \rmd^D\bm{k}_1
\int \rmd^D\bm{k}_2
\int \rmd^D\bm{k}_3\,
K_{\bm{k}_1\bm{k}_2\bm{k}_3}J_{\bm{k}_1}J_{\bm{k}_2}J_{\bm{k}_3}^*
\nonumber\\
&\qquad\qquad\qquad\qquad\qquad\qquad\qquad\qquad\qquad\qquad\quad\ \ %
{}+{}\cdots{}
\label{eqn:3rdCumulant}
\end{align}
in the cumulant expansion of $\ln\chi_0[J,J^*]$.
As we did for the second-order correlations, we allow this third-order correlation $K_{\bm{k}_1\bm{k}_2\bm{k}_3}$ to possess a translationally invariant component proportional to $\delta^D(\bm{k}_1+\bm{k}_2-\bm{k}_3)$; otherwise, it is assumed to be free from singularity.
Namely, it assumes the form
\begin{equation}
K_{\bm{k}_1\bm{k}_2\bm{k}_3}
=\bar{K}_{\bm{k}_1\bm{k}_2}
\delta^D(\bm{k}_1+\bm{k}_2-\bm{k}_3)
+\tilde{K}_{\bm{k}_1\bm{k}_2\bm{k}_3},
\label{eqn:3rdCumulantStruc}
\end{equation}
with $\bar{K}_{\bm{k}_1\bm{k}_2}$ and $\tilde{K}_{\bm{k}_1\bm{k}_2\bm{k}_3}$ being regular functions of the momenta.
It physically means that this third-order correlation is allowed to exist everywhere in space, while its correlation lengths are finite. 
This cumulant evolves in time according to the Hamiltonian (\ref{eqn:FreeBoson}) as
\begin{align}
&\ln\chi_t[J,J^*]
\nonumber
\displaybreak[0]
\\
&\ %
={}\cdots{}
+\int \rmd^D\bm{k}_1
\int \rmd^D\bm{k}_2
\int \rmd^D\bm{k}_3\,
K_{\bm{k}_1\bm{k}_2\bm{k}_3}J_{\bm{k}_1}J_{\bm{k}_2}J_{\bm{k}_3}^*
\nonumber\\[-0.8truemm]
&\qquad\qquad\qquad\qquad\qquad\qquad\qquad\quad\ %
{}
\times
\rme^{\rmi(\omega_{k_1}+\omega_{k_2}-\omega_{k_3})t}
\nonumber
%\displaybreak[0]
\\
&\qquad\qquad\qquad\qquad\qquad\qquad\qquad\qquad\qquad\qquad\quad
{}
+{}\cdots{}.
\label{eqn:3rdCumulantEvo}
\end{align}
This decays in the long-time limit $t\to\infty$ for a generic dispersion relation $\omega_k$, according to the Riemann-Lebesgue lemma.
In this way, any higher-order cumulants decay in the long-time limit $t\to\infty$ under the assumption of finite correlation lengths mentioned above (they are regular apart from the translationally invariant components), and the system relaxes to the Gaussian GGE as (\ref{Characteristic2}), even from a non-Gaussian initial state.

See Appendix \ref{app:DecayCumulants}, where explicit formulas for the time evolutions of all the cumulants for Gaussian spectra with a quadratic dispersion relation $\omega_k$ are provided.
All the cumulants except for the translationally invariant component of the second-order normal correlation decay in the long-time limit $t\to\infty$.
The results for the Gaussian spectra are of quite general validity, since by the saddle-point approximation the integrals for the cumulants are approximated by Gaussian integrals for large $t$.
For any quadratic dispersion relation $\omega_k$, the cumulants exhibit power-law decays as shown in (\ref{eqn:AsympDecayLocal}) and (\ref{eqn:AsympDecayNonLocal}).

If the third- and higher-order correlations exist only locally, with no translationally invariant components, we can show the relaxation to the Gaussian GGE in a simpler way as follows.
First, we observe that the Gaussian GGE in (\ref{density1}) is \textit{mixing} with respect to the Hamiltonian (\ref{eqn:FreeBoson}) \cite{Bratteli1}, namely,
\begin{equation}
\langle \hat{A}\hat{B}(t)\hat{C}\rangle_\text{GGE}
\xrightarrow{t\to\infty}
\langle\hat{A}\hat{C}\rangle_\text{GGE}\langle\hat{B}\rangle_\text{GGE}
\label{mixing1}
\end{equation}
holds for any local observables $\hat{A}$, $\hat{B}$, and $\hat{C}$, where $\hat{B}(t)=\rme^{\rmi\hat{H}t}\hat{B}\rme^{-\rmi\hat{H}t}$ is the Heisenberg operator corresponding to $\hat{B}$, and $\langle{}\cdots{}\rangle_\text{GGE}$ denotes the expectation value in the Gaussian GGE in (\ref{density1}). See Appendix \ref{app:Mixing} for a proof.
This tells us that any locally perturbed GGE, which is non-Gaussian in general, returns to the Gaussian GGE \cite{ref:ReturnEq}:
\begin{align}
	\hat{\rho}_0
	&=\sum_j\hat{L}_j\hat{\rho}_\text{GGE}\hat{L}_j^\dag
	\nonumber\\
	&\mapsto\hat{\rho}(t)= \rme^{-\rmi\hat{H}t} \hat{\rho}_0\rme^{\rmi\hat{H}t}\xrightarrow{t\to\infty}\hat{\rho}_\text{GGE},
	\label{eqn:return}
\end{align}
where $\hat{L}_j$ represent the local perturbations, satisfying $\sum_j\hat{L}_j^\dag\hat{L}_j=\hat{\openone}$ for the normalization of $\hat{\rho}_0$.
Indeed, thanks to the mixing property (\ref{mixing1}), we get
\begin{align}
\langle\hat{A}\rangle_t
&=\langle\hat{A}(t)\rangle
\vphantom{\sum_i}
\nonumber\displaybreak[0]\\
&=\sum_j\langle\hat{L}_j^\dag\hat{A}(t)\hat{L}_j\rangle_\text{GGE}
\nonumber\displaybreak[0]\\
&\xrightarrow{t\to\infty}
\sum_j\langle\hat{L}_j^\dag\hat{L}_j\rangle_\text{GGE}\langle\hat{A}\rangle_\text{GGE}
\nonumber\displaybreak[0]\\
&=\langle\hat{A}\rangle_\text{GGE}
\end{align}
for any local observable $\hat{A}$, where $\langle{}\cdots{}\rangle$ denotes the expectation value in the initial state $\hat{\rho}_0$ in (\ref{eqn:return}).
This proves the relaxation to the Gaussian GGE in (\ref{eqn:return}).

If the third-order cumulant $K_{\bm{k}_1\bm{k}_2\bm{k}_3}$ in (\ref{eqn:3rdCumulant}) in the initial state contains a contribution proportional to $\delta(\omega_{k_1}+\omega_{k_2}-\omega_{k_3})$, it is stationary and survives in the long-time limit $t\to\infty$.
This adds integrals of motion like $\delta(\omega_{k_1}+\omega_{k_2}-\omega_{k_3})\hat{b}_{\bm{k}_1}\hat{b}_{\bm{k}_2}\hat{b}_{\bm{k}_3}^\dag$ to the exponent of the GGE in (\ref{density1}).
If, however, the initial state is prepared irrespective of the Hamiltonian $\hat{H}$ of the system, e.g.\ in a quench scenario, it is typically impossible that the initial state is equipped with such a delicate structure in the correlations, matching the characteristics of the Hamiltonian $\hat{H}$.
The same applies to the higher-order cumulants.
The relaxation to the Gaussian GGE in (\ref{density1}) is therefore quite general for generic initial states.

\section{Harmonic Oscillator Coupled with Bosonic Reservoirs}
\label{sec:Friedrichs}
In the previous section, we have observed the relaxation of the free bosonic field to a Gaussian GGE\@.
Let us now look at a coupled system.
As an interesting and nontrivial example, we consider a harmonic oscillator coupled with multiple bosonic reservoirs.
The Hamiltonian is given by \cite{ref:OscillatorFriedrichsBarnett,ref:ExactDiagonalizationCaldeira,ref:FactorizeModel,ref:FTExactMonnai,Monnai1}
\begin{equation}
	\hat{H}=\hat{H}_S+\sum_\ell\hat{H}_\ell+\lambda\hat{V}
\label{Hamiltonian2}
\end{equation}
with
\begin{align}
&\hat{H}_S=
\Omega \hat{a}^\dag \hat{a},\qquad
\hat{H}_\ell
=\int \rmd^D\bm{k}\,\omega_{k\ell}\hat{b}_{\bm{k}\ell}^\dag \hat{b}_{\bm{k}\ell},
%\displaybreak[0]
\\
&\hat{V}
=\sum_\ell\int \rmd^D\bm{k}\,(u_{\bm{k}\ell}^*\hat{a}^\dag \hat{b}_{\bm{k}\ell}+u_{\bm{k}\ell}\hat{b}_{\bm{k}\ell}^\dag \hat{a}).
\end{align}
The operators $\hat{a}$ and $\hat{b}_{\bm{k}\ell}$ are the canonical operators for the harmonic oscillator and the bosonic reservoirs, respectively, satisfying the canonical commutation relations
\begin{gather}
[\hat{a},\hat{a}]=0,\quad
[\hat{a},\hat{a}^\dag]=1,
%\displaybreak[0]
\\
[\hat{b}_{\bm{k}\ell},\hat{b}_{\bm{k}'\ell'}]=0,\quad
[\hat{b}_{\bm{k}\ell},\hat{b}_{\bm{k}'\ell'}^\dag]=\delta_{\ell\ell'}\delta^D(\bm{k}-\bm{k}'),
%\displaybreak[0]
\\
[\hat{a},\hat{b}_{\bm{k}\ell}]
=[\hat{a},b_{\bm{k}\ell}^\dag]=0.
\end{gather}
$\Omega>0$ is the frequency of the harmonic oscillator, $\omega_{k\ell}\ge0$ is the dispersion relation for the $\ell$th reservoir, $u_{\bm{k}\ell}$ is the form factor of the interaction between the harmonic oscillator and the $\ell$th reservoir \cite{note:FormFactor}, and $\lambda$ is a coupling constant \cite{note:BoundState}. %\cite{ref:Miyamoto-BoundStatesPRA2005}.

It is an exactly solvable quadratic model, and we can diagonalize the Hamiltonian (\ref{Hamiltonian2}) as \cite{ref:OscillatorFriedrichsBarnett,ref:ExactDiagonalizationCaldeira,ref:FTExactMonnai,ref:FanoDiagonalization}
\begin{equation}
\hat{H}=\sum_\ell\int \rmd^D\bm{k}\,\omega_{k\ell}\hat{A}_ {\bm{k}\ell}^\dagger\hat{A}_ {\bm{k}\ell},
\label{diagonal1}
\end{equation}
where the normal modes are given by
\begin{equation}
\hat{A}_{\bm{k}\ell}
=
\alpha_{\bm{k}\ell}
\hat{a}
+\sum_{\ell'}\int \rmd^D\bm{k}'\,\beta_{\bm{k}\ell,\bm{k}'\ell'}\hat{b}_{\bm{k}'\ell'},
\label{eqn:NormalModes}
\end{equation}
with
\begin{align}
\alpha_{\bm{k}\ell}
&=
\frac{\lambda u_{\bm{k}\ell}}{\omega_{k\ell}-\Omega-\lambda^2\Sigma(\omega_{k\ell}-\rmi0^+)},
\displaybreak[0]\\
\beta_{\bm{k}\ell,\bm{k}'\ell'}
&=\delta_{\ell\ell'}\delta^D(\bm{k}-\bm{k}')
+
\frac{\lambda\alpha_{\bm{k}\ell}u_{\bm{k}'\ell'}^*}{\omega_{k\ell}-\omega_{k'\ell'}-\rmi0^+}.
\end{align}
Here, the self-energy function on the complex $z$ plane is given by
\begin{equation}
\Sigma(z)=\sum_\ell\int \rmd^D\bm{k}\,\frac{|u_{\bm{k}\ell}|^2}{z-\omega_{k\ell}},
\end{equation}
and the coefficients satisfy the orthogonality
\begin{gather}
\sum_\ell\int\rmd^D\bm{k}\,|\alpha_{\bm{k}\ell}|^2=1,\quad
\sum_\ell\int\rmd^D\bm{k}\,\alpha_{\bm{k}\ell}^*\beta_{\bm{k}\ell,\bm{k}'\ell'}=0,
\displaybreak[0]
\label{eqn:Orthogonality1}
\\
\sum_\ell\int\rmd^D\bm{k}\,\beta_{\bm{k}\ell,\bm{k}'\ell'}^*\beta_{\bm{k}\ell,\bm{k}''\ell''}=\delta_{\ell'\ell''}\delta^D(\bm{k}'-\bm{k}''),
\label{eqn:Orthogonality2}
\end{gather}
and the completeness
\begin{equation}
\alpha_{\bm{k}\ell}
\alpha_{\bm{k}'\ell'}^*
+\sum_{\ell''}\int\rmd^D\bm{k}''\,\beta_{\bm{k}\ell,\bm{k}''\ell''}\beta_{\bm{k}'\ell',\bm{k}''\ell''}^*=\delta_{\ell\ell'}\delta^D(\bm{k}-\bm{k}').
\label{eqn:Completeness}
\end{equation}
The completeness (\ref{eqn:Completeness}) ensures that the operators $\hat{A}_{\bm{k}\ell}$ satisfy the canonical commutation relations
\begin{equation}
[\hat{A}_{\bm{k}\ell},\hat{A}_{\bm{k}'\ell'}]=0,\quad
[\hat{A}_{\bm{k}\ell},\hat{A}_{\bm{k}'\ell'}^\dag]=\delta_{\ell\ell'}\delta^D(\bm{k}-\bm{k}'),
\end{equation}
and the orthogonality (\ref{eqn:Orthogonality1})--(\ref{eqn:Orthogonality2}) allows us to invert the relation (\ref{eqn:NormalModes}) as
\begin{equation}
	\hat{a}=\sum_{\ell}\int\rmd^D\bm{k}\,\alpha_{\bm{k}\ell}^*\hat{A}_{\bm{k}\ell},\ \ %
	\hat{b}_{\bm{k}\ell}
	=\sum_{\ell'}\int\rmd^D\bm{k}'\,\beta_{\bm{k}'\ell',\bm{k}\ell}^*\hat{A}_{\bm{k}'\ell'}.
	\label{eqn:Inversion}
\end{equation}

\subsection{Factorized Initial State}
\label{sec:FactorizedInitialState}
We first consider the evolution of the system from a factorized initial state
\begin{equation}
\hat{\rho}_0=\hat{\rho}_S\otimes\biggl(\mathop{\bigotimes}_\ell\hat{\rho}_\ell\biggr),
\label{eqn:InitialProductState}
\end{equation}
with no correlations among the harmonic oscillator $\hat{\rho}_S$ and the reservoirs $\hat{\rho}_\ell$.
The characteristic function of the initial state of the harmonic oscillator $\hat{\rho}_S$ and the characteristic functional of the initial state of the reservoirs $\hat{\rho}_B=\mathop{\bigotimes}_\ell\hat{\rho}_\ell$ are defined respectively by
\begin{align}
\chi_S(\xi,\xi^*)
&=
\langle
\rme^{\xi\hat{a}^\dag-\xi^*\hat{a}}
\rangle,
\label{eqn:CharFuncS}
\displaybreak[0]
\\
\chi_B[\eta,\eta^*]
&=
\langle
\rme^{\sum_\ell\int\rmd^D\bm{k}\,(\eta_{\bm{k}\ell}\hat{b}_{\bm{k}\ell}^\dag-\eta_{\bm{k}\ell}^*\hat{b}_{\bm{k}\ell})}
\rangle,
\label{eqn:CharFuncB}
\end{align}
where $\langle{}\cdots{}\rangle$ denotes the expectation value in the initial state $\hat{\rho}_0$ in (\ref{eqn:InitialProductState}).
The initial state of the harmonic oscillator $\hat{\rho}_S$ is arbitrary, while the initial state of each bosonic reservoir is a (non-Gaussian) state like the one considered in Sec.\ \ref{sec:FreeBoson}, under the assumption of finite correlation lengths.
More specifically, in the cumulant expansion of the characteristic functional of the reservoirs $\chi_B[\eta,\eta^*]$,
\begin{align}
&\ln\chi_B[\eta,\eta^*]
\vphantom{\int}
\nonumber\\
&\ %
=\sum_\ell
\int\rmd^D\bm{k}\,(\eta_{\bm{k}\ell}\langle\hat{b}_{\bm{k}\ell}^\dag\rangle-\eta_{\bm{k}\ell}^*\langle\hat{b}_{\bm{k}\ell}\rangle)
\nonumber\\
&\ \quad
{}-\frac{1}{2}\sum_\ell\int\rmd^D\bm{k}\int\rmd^D\bm{k}'
	\begin{pmatrix}
\eta_{\bm{k}\ell}^*
&
\eta_{\bm{k}\ell}
\end{pmatrix}
V_{\bm{k}\bm{k}'}^{(\ell)}
\begin{pmatrix}
\medskip
\eta_{\bm{k}'\ell}
\\
\eta_{\bm{k}'\ell}^*
\end{pmatrix}
\nonumber\\
&\ \quad
{}+\sum_\ell
\int \rmd^D\bm{k}_1
\int \rmd^D\bm{k}_2
\int \rmd^D\bm{k}_3\,
K_{\bm{k}_1\bm{k}_2\bm{k}_3}^{(\ell)}
\eta_{\bm{k}_1\ell}\eta_{\bm{k}_2\ell}\eta_{\bm{k}_3\ell}^*
\nonumber\\
&\ \quad
{}+{}\cdots{},
\vphantom{\sum}
\label{eqn:CumExpB}
\end{align}
the covariance matrix $V_{\bm{k}\bm{k}'}^{(\ell)}$ and the third-order cumulant $K_{\bm{k}_1\bm{k}_2\bm{k}_3}^{(\ell)}$ of the $\ell$th reservoir are endowed with the same structures as those in (\ref{eqn:CovMatStruc}) and (\ref{eqn:3rdCumulantStruc}), respectively, with $f_{\bm{k}}$, $F_{\bm{k}\bm{k}'}$, $g_{\bm{k}}$, $G_{\bm{k}\bm{k}'}$, $\bar{K}_{\bm{k}_1\bm{k}_2}$, and $\tilde{K}_{\bm{k}_1\bm{k}_2\bm{k}_3}$ replaced by $f_{\bm{k}}^{(\ell)}$, $F_{\bm{k}\bm{k}'}^{(\ell)}$, $g_{\bm{k}}^{(\ell)}$, $G_{\bm{k}\bm{k}'}^{(\ell)}$, $\bar{K}_{\bm{k}_1\bm{k}_2}^{(\ell)}$, and $\tilde{K}_{\bm{k}_1\bm{k}_2\bm{k}_3}^{(\ell)}$, which are all assumed to be regular functions of the momenta.
The other (higher-order) cumulants of the reservoirs can also possess translationally invariant components, but otherwise they are assumed to be regular in momenta.

Starting from such a factorized initial state $\hat{\rho}_0$ in (\ref{eqn:InitialProductState}), the characteristic functional of the state of the total system evolves according to the Hamiltonian (\ref{diagonal1}) as
\begin{align}
\chi_t[J,J^*]
&=
\langle
\rme^{\sum_\ell\int\rmd^D\bm{k}\,(J_{\bm{k}\ell}\hat{A}_{\bm{k}\ell}^\dag-J_{\bm{k}\ell}^*\hat{A}_{\bm{k}\ell})}
\rangle_t
\nonumber
\displaybreak[0]
\\
&=
\langle
\rme^{\sum_\ell\int\rmd^D\bm{k}\,(J_{\bm{k}\ell}\hat{A}_{\bm{k}\ell}^\dag\rme^{\rmi\omega_{k\ell}t}-J_{\bm{k}\ell}^*\hat{A}_{\bm{k}\ell}\rme^{-\rmi\omega_{k\ell}t})}
\rangle
\nonumber
\displaybreak[0]
\\
&=
\chi_S\bm{(}\xi(t),\xi^*(t)\bm{)}
\chi_B[\eta(t),\eta^*(t)],
\label{eqn:CharFuncTot}
\end{align}
where $\chi_S(\xi,\xi^*)$ and $\chi_B[\eta,\eta^*]$ are the characteristic function and the characteristic functional of the initial states of the harmonic oscillator and of the reservoirs given respectively in (\ref{eqn:CharFuncS}) and (\ref{eqn:CharFuncB}), and 
\begin{align}
	\xi(t)
	&=\sum_\ell\int\rmd^D\bm{k}\,\alpha_{\bm{k}\ell}^*
	\rme^{\rmi\omega_{k\ell}t}
	J_{\bm{k}\ell},
	\displaybreak[0]\\
	\eta_{\bm{k}\ell}(t)
	&=\sum_{\ell'}\int \rmd^D\bm{k}'\,
	\beta_{\bm{k}'\ell',\bm{k}\ell}^*
	\rme^{\rmi\omega_{k'\ell'}t}
	J_{\bm{k}'\ell'}.
\end{align}
In the long-time limit $t\to\infty$, we get
\begin{widetext}
\begin{equation}
	\xi(t)
	\xrightarrow{t\to\infty}0
	\label{eqn:XiDecay}
\end{equation}
due to the Riemann-Lebesgue lemma, while
\begin{equation}
	\eta_{\bm{k}\ell}(t)\rme^{-\rmi\omega_{k\ell}t}
	=\sum_{\ell'}\int \rmd^D\bm{k}'\left(
	\delta_{\ell\ell'}\delta^D(\bm{k}-\bm{k}')
-
\frac{\lambda u_{\bm{k}\ell}\alpha_{\bm{k}'\ell'}^*}{\omega_{k\ell}-\omega_{k'\ell'}-\rmi0^+}
\right)
	\rme^{-\rmi(\omega_{k\ell}-\omega_{k'\ell'})t}
	J_{\bm{k}'\ell'}
\xrightarrow{t\to\infty}J_{\bm{k}\ell}
	\label{eqn:EtaDecay}
\end{equation}
\end{widetext}
by recalling
\begin{equation}
\frac{\rme^{-\rmi\omega t}}{\omega-\rmi0^+}
\to
\begin{cases}
\medskip
0&(t\to+\infty),\\
2\pi\rmi\delta(\omega)&(t\to-\infty).
\end{cases}
\end{equation}
Therefore, the characteristic functional of the total system (\ref{eqn:CharFuncTot}) behaves asymptotically as
\begin{equation}
\chi_t[J,J^*]
\xrightarrow{t\to\infty}
\chi_B[J\rme^{\rmi\omega t},J^*\rme^{-\rmi\omega t}],
\label{eqn:CharFuncTotAsymp}
\end{equation}
where we have used the normalization condition $\chi_S(0,0)=1$ of the initial state $\hat{\rho}_S$ of the harmonic oscillator.
Notice that the asymptotic characteristic functional $\chi_B[J\rme^{\rmi\omega t},J^*\rme^{-\rmi\omega t}]$ in (\ref{eqn:CharFuncTotAsymp}) is essentially the same as (\ref{Characteristic1}), and the results in Sec.\ \ref{sec:FreeBoson} immediately apply.
Each reservoir relaxes as (\ref{Characteristic2}), and the characteristic functional of the total system in (\ref{eqn:CharFuncTotAsymp}) further relaxes to
\begin{align}
\chi_t[J,J^*]
&\xrightarrow{t\to\infty}
\exp\!\left(
-\frac{1}{2}
\sum_\ell
\int \rmd^D\bm{k}\,
(1+2f_{\bm{k}}^{(\ell)})|J_{\bm{k}\ell}|^2
\right)
\nonumber
\displaybreak[0]
\\
&
=\chi_\text{NESS}[J,J^*].
\label{eqn:GNESS}
\end{align}
In terms of the density operator, the stationary state is formally given by
\begin{equation}
\hat{\rho}_\text{NESS}
\propto
\exp\!\left(
-\sum_\ell
\int \rmd^D\bm{k}
\ln(1+f_{\bm{k}}^{(\ell){-1}})
\hat{A}_{\bm{k}\ell}^\dag
\hat{A}_{\bm{k}\ell}
\right).
\label{eqn:GNESSop}
\end{equation}
It is a Gaussian state: the total system relaxes to the Gaussian state $\hat{\rho}_\text{NESS}$, even from a non-Gaussian initial state $\hat{\rho}_0$, under the condition of finite correlation lengths in the reservoirs.

Let us comment on some physical aspects of the stationary state $\hat{\rho}_\text{NESS}$.

\subsubsection{Nonequilibrium Steady State}
In the presence of two or more reservoirs, the stationary state (\ref{eqn:GNESSop}) is a nonequilibrium steady state (NESS) \cite{ref:NESS-AschbacherPillet-JSP,ref:NESS-AschbacherJaksicPautratPillet,ref:NESS-Tasaki,ref:FTExactMonnai,Monnai1}, in which steady currents flow among the reservoirs through the harmonic oscillator.
That is why we have named the stationary state $\hat{\rho}_\text{NESS}$ in (\ref{eqn:GNESSop}).
For instance, the energy current
\begin{equation}
\hat{J}_\ell=-\rmi[\hat{H}_\ell,\hat{H}]
=\rmi\lambda\int\rmd^D\bm{k}\,\omega_{k\ell}(
u_{\bm{k}\ell}^*\hat{a}^\dag\hat{b}_{\bm{k}\ell}
-u_{\bm{k}\ell}\hat{b}_{\bm{k}\ell}^\dag\hat{a}
)
\end{equation}
flowing into the $\ell$th reservoir per time is estimated in the stationary state $\hat{\rho}_\text{NESS}$ to be
\begin{align}
&\langle\hat{J}_\ell\rangle_\text{NESS}
\nonumber\\
&\ \ %
=-2\lambda\Im
\sum_{\ell'}
\int\rmd^D\bm{k}
\int\rmd^D\bm{k}'\,
\omega_{k\ell}
u_{\bm{k}\ell}^*
\beta_{\bm{k}'\ell',\bm{k}\ell}^*
f_{\bm{k}'}^{(\ell')}
\alpha_{\bm{k}'\ell'}
\nonumber\displaybreak[0]\\
&\ \ %
=-\lambda^2
\int_0^\infty\rmd\omega\,
\omega
|\alpha(\omega)|^2
\left(
\mathcal{F}_\ell(\omega)
-
\mathcal{F}(\omega)
\frac{
\Gamma_\ell(\omega)
}{\Gamma(\omega)}
\right),
\label{eqn:Current}
\end{align}
where
\begin{align}
\Gamma(\omega)
&=\sum_\ell\Gamma_\ell(\omega),\quad
\mathcal{F}(\omega)
=\sum_\ell\mathcal{F}_\ell(\omega),\\%
\Gamma_\ell(\omega)
&=2\pi\int\rmd^D\bm{k}\,|u_{\bm{k}\ell}|^2\delta(\omega_{k\ell}-\omega),
\\
\mathcal{F}_\ell(\omega)
&=2\pi\int\rmd^D\bm{k}\,f_{\bm{k}}^{(\ell)}|u_{\bm{k}\ell}|^2\delta(\omega_{k\ell}-\omega),
\end{align}
and we have introduced
\begin{equation}
\alpha(\omega)
=\frac{\lambda\sqrt{\Gamma(\omega)/2\pi}}{\omega-\Omega-\lambda^2\Sigma(\omega-\rmi0^+)},
\end{equation}
which is normalized as $\int_0^\infty\rmd\omega\,|\alpha(\omega)|^2=1$, and $|\alpha(\omega)|^2\to\delta(\omega-\Omega)$ in the weak-coupling limit $\lambda\to0$.
If $f_{\bm{k}}^{(\ell)}$ is isotropic and is given by $f_{\bm{k}}^{(\ell)}=f_\ell(\omega_{k\ell})$ [e.g., in the case of the canonical ensemble, $f_\ell(\omega)$ is the Bose distribution function], we have $\mathcal{F}_\ell(\omega)=f_\ell(\omega)\Gamma_\ell(\omega)$, and the current (\ref{eqn:Current}) is simplified to
\begin{align}
&\langle\hat{J}_\ell\rangle_\text{NESS}
\nonumber\\
&\,
=-\lambda^2
\sum_{\ell'}
\int_0^\infty\rmd\omega\,
\omega
|\alpha(\omega)|^2
[
f_\ell(\omega)-f_{\ell'}(\omega)
]
\frac{
\Gamma_\ell(\omega)
\Gamma_{\ell'}(\omega)
}{\Gamma(\omega)}.
\label{eqn:fomega}
\end{align}
The steady current flows by the difference between $f_\ell(\omega)$ and $f_{\ell'}(\omega)$.
In the weak-coupling limit $\lambda\to0$ (more precisely, in the van Hove limit $\lambda\to0$ keeping the scaled time $\tau=\lambda^2t$ finite), it is further simplified to 
\begin{equation}
\frac{1}{\lambda^2}\langle\hat{J}_\ell\rangle_\text{NESS}
\xrightarrow{\lambda\to0}
-
\Omega
\sum_{\ell'}
[
f_\ell(\Omega)-f_{\ell'}(\Omega)
]
\frac{
\Gamma_\ell(\Omega)
\Gamma_{\ell'}(\Omega)
}{\Gamma(\Omega)},
\label{eqn:fomegaWeak}
\end{equation}
which is a standard Landauer formula but with the reservoirs in GGEs.

\subsubsection{Equilibration of the Subsystem}
The long-time limit (\ref {eqn:GNESS}) shows that the system forgets the initial state $\hat{\rho}_S$ of the harmonic oscillator and relaxes to the stationary state $\hat{\rho}_\text{NESS}$ independent of $\hat{\rho}_S$.
Let us look in which state the harmonic oscillator equilibrates.
Recalling the inversion formula (\ref{eqn:Inversion}), the characteristic function of the harmonic oscillator $\chi_t^{(S)}(\xi,\xi^*)$ can be extracted from the characteristic functional of the total system $\chi_t[J,J^*]$ in (\ref{eqn:CharFuncTot}).
It relaxes to
\begin{align}
\chi_t^{(S)}(\xi,\xi^*)
&=\langle\rme^{\xi\hat{a}^\dag-\xi^*\hat{a}}\rangle_t
\nonumber\displaybreak[0]\\
&=\chi_t[\alpha\xi,\xi^*\alpha^*]
\vphantom{\int}
\nonumber\displaybreak[0]\\
&\xrightarrow{t\to\infty}
\chi_\text{NESS}[\alpha\xi,\xi^*\alpha^*]
\vphantom{\int}
\nonumber\displaybreak[0]\\
&=
\exp\!\left(
-\frac{1}{2}
|\xi|^2
\sum_\ell
\int \rmd^D\bm{k}\,
|\alpha_{\bm{k}\ell}|^2
(1+2f_{\bm{k}}^{(\ell)})
\right)
\nonumber\displaybreak[0]\\
&=
\exp\!\left[
-\frac{1}{2}
|\xi|^2
\int_0^\infty\rmd\omega\,
|\alpha(\omega)|^2
\left(
1+2\frac{\mathcal{F}(\omega)}{\Gamma(\omega)}
\right)
\right]
\nonumber\displaybreak[0]\\
&=\chi_\text{NESS}^{(S)}(\xi,\xi^*).
\vphantom{\int}
\end{align}
The corresponding equilibrium density operator is given by
\begin{equation}
\hat{\rho}_\text{NESS}^{(S)}
\propto
\rme^{-\theta\hat{a}^\dag\hat{a}}
\label{eqn:SNESS}
\end{equation}
with
\begin{equation}
\theta
=2
\coth^{-1}\!\left[
\int_0^\infty\rmd\omega\,
|\alpha(\omega)|^2
\left(
1+2\frac{\mathcal{F}(\omega)}{\Gamma(\omega)}
\right)
\right].
\end{equation}
If the harmonic oscillator is immersed in a single reservoir and $f_{\bm{k}}=f(\omega_k)=1/(\rme^{\omega_k/k_BT}-1)$ is the Bose distribution function, the equilibrium density operator $\hat{\rho}_\text{NESS}^{(S)}$ in (\ref{eqn:SNESS}) is reduced to the thermal state $\hat{\rho}_\text{NESS}^{(S)}\propto\rme^{-\hat{H}_S/k_BT}$ at the same temperature $T$ as that of the reservoir in the weak-coupling limit $\lambda\to0$.
In general, the reservoirs do not relax to the canonical state, with $f_{\bm{k}}^{(\ell)}$ being different from the Bose distribution functions, but in any case the equilibrium state of the harmonic oscillator $\hat{\rho}_\text{NESS}^{(S)}$ in (\ref{eqn:SNESS}) looks like a canonical state with an effective temperature $\Omega/k_B\theta$, which depends on the GGE characterized by $f_{\bm{k}}^{(\ell)}$.

\subsection{Correlated Initial State}
In the previous subsection, we considered factorized initial states (\ref{eqn:InitialProductState}), with no correlations among the harmonic oscillator and the reservoirs.
Even if there are some correlations in the initial state, the system relaxes to the same stationary state $\hat{\rho}_\text{NESS}$ as that given in (\ref{eqn:GNESSop}), as long as the initial correlations are just local.
Namely, we consider, instead of the factorized initial state (\ref{eqn:InitialProductState}), a correlated initial state
\begin{align}
\hat{\rho}_0
=\sum_j\hat{L}_j
\,\biggl[
\hat{\rho}_S\otimes\biggl(\mathop{\bigotimes}_\ell\hat{\rho}_\ell\biggr)
\biggr]\,
\hat{L}_j^\dag,
\label{eqn:CorrelatedInitialState}
\end{align}
where the factorized state (\ref{eqn:InitialProductState}) is perturbed by local operators $\hat{L}_j$, which satisfy $\sum_j\hat{L}_j^\dag\hat{L}_j=\openone$ and induce correlations among the harmonic oscillator and the reservoirs.
Still, the system relaxes to
\begin{equation}
\hat{\rho}(t)= \rme^{-\rmi\hat{H}t} \hat{\rho}_0\rme^{\rmi\hat{H}t}\xrightarrow{t\to\infty}\hat{\rho}_\text{NESS},
\label{eqn:RelaxCorrelated}
\end{equation}
where $\hat{\rho}_\text{NESS}$ is the same stationary state as the one presented in (\ref{eqn:GNESSop}).
A proof is provided in Appendix \ref{app:RelaxCorrelated}\@.

\section{Summary}
\label{sec:Conclusions}
We have shown a scenario of the relaxation to GGE for integrable models which can be mapped to free bosonic fields in the thermodynamic limit. 
The unitary transformation to the free bosonic fields would be regarded as a quantum counterpart of the canonical transformation to an assembly of harmonic oscillators for classical systems in the context of the Liouville integrability.
Then, the field operators in the characteristic functional just acquire oscillating factors in the evolution as in (\ref{Characteristic1}), and the diagonal singularity yields the stationary state. 
As a result, only the gauge invariant terms such as the occupation number $\hat{I}_{\bm{k}}=\hat{b}_{\bm{k}}^\dag\hat{b}_{\bm{k}}$ survive in the GGE in (\ref{Characteristic2}) in the long-time limit $t\rightarrow\infty$.

Moreover, the GGE (\ref{Characteristic2}) is a simple Gaussian state: only the quadratic gauge invariant terms $\hat{I}_{\bm{k}}=\hat{b}_{\bm{k}}^\dag\hat{b}_{\bm{k}}$ contribute to the GGE, as long as the initial (not necessarily Gaussian) state fulfills a few physical conditions, where the presence of anomalous correlations is allowed.
In contrast to the previous works \cite{ref:CramersDawsonEisertOsbornePRL2008,ref:CramerEisert-CentralLimitTheorem,ref:GogolinEisert-GaussificationFermi,Farrelly1,Murthy1}, which proved the Gaussification for large but finite systems, we have directly analyzed the systems in the thermodynamic limit, with continuous spectra.
This greatly simplifies the analysis, and in this picture, the mechanism for the equilibration is due to the Riemann-Lebesgue lemma and the diagonal singularity. 
We have solved the evolutions of the states of the whole systems exactly, and have shown the Gaussification for rather general (non-Gaussian) initial states (particles can be distributed all over the space and correlations can be present everywhere in space, but the correlations should be of finite range), even in the presence of initial correlations among systems.

We stress that the observables are also relevant to the relaxation.
In particular, the locality of the observables is important, for the Riemann-Lebesgue lemma to work.
It is known that the relaxation times for typical systems and/or typical settings are typically short \cite{relaxation1,Goldstein3,Monnai2,relaxation2,Goldstein1b,relaxation3} irrespective of the system characteristics or the observables.
For specific (atypical) systems, however, it is not the case, and the relaxation time depends on the choice of the observable as well as the initial state.

In this work, we have studied the relaxation to a GGE for time-independent Hamiltonians.
It would be interesting, as a potential future subject, to extend the discussion to the stationary states of periodically driven systems, on the basis of the Floquet theory \cite{Lazarides1,Lazarides2,DAlessio1,Haldar1}.

\begin{acknowledgments}
This work was supported by the Grants-in-Aid for Scientific Research (C) (No.~18K03467 and No.~18K03470) and for Fostering Joint International Research (B) (No.~18KK0073) both from the Japan Society for the Promotion of Science (JSPS), and by the Waseda University Grant for Special Research Projects (No.~2018K-262).
\end{acknowledgments}

\appendix
\section{Decay of Correlations}
\label{app:DecayCumulants}
The Gaussification is a consequence of the decay of the cumulants except for the occupation numbers.
We have attributed the mechanism behind the decay to the Riemann-Lebesgue lemma, under the assumptions (i)--(v) listed in Sec.\ \ref{sec:Introduction}\@.
Let us here compute the evolutions of the cumulants explicitly for analytically tractable specific forms of spectra to get an idea on how the Riemann-Lebesgue lemma works and how the correlations decay in the long-time limit.

Let us first look at the second-order cumulants in (\ref{eqn:GaussCharEvolve}).
There are four different types.
The translationally invariant component of the normal correlation represented by $f_{\bm{k}}$, which is related to the occupation numbers, does not decay and survives in the GGE: $\bar{\mathcal{K}}_{1,1}(t)=\int\rmd^D\bm{k}\,f_{\bm{k}}J_{\bm{k}}J_{\bm{k}}^*$ in (\ref{eqn:GaussCharEvolve}) is independent of time $t$.
The remaining (local) component of the normal correlation represented by $F_{\bm{k}\bm{k}'}$ evolves in time as $\tilde{\mathcal{K}}_{1,1}(t)=\int\rmd^D\bm{k}\int\rmd^D\bm{k}'\,F_{\bm{k}\bm{k}'}J_{\bm{k}}J_{\bm{k}'}^*\rme^{\rmi(\omega_k-\omega_{k'})t}$.
In general, the cumulant involving $m_+$ creation operators $\hat{b}_{\bm{k}}^\dag$ and $m_-$ annihilation operators $\hat{b}_{\bm{k}}$ evolves in time as
\begin{widetext}
\begin{align}
\mathcal{K}_{m_+,m_-}(t)
&=
\int \rmd^D\bm{k}_1
\cdots
\int \rmd^D\bm{k}_{m_+}
\int \rmd^D\bm{k}_1'
\cdots
\int \rmd^D\bm{k}_{m_-}'
K_{\bm{k}_1\ldots\bm{k}_{m_+}\bm{k}_1'\ldots\bm{k}_{m_-}'}
J_{\bm{k}_1}\cdots J_{\bm{k}_{m_+}}
J_{\bm{k}_1'}^*\cdots J_{\bm{k}_{m_-}'}^*
\nonumber\\[-1truemm]
&\qquad\qquad\qquad\qquad\qquad\qquad\qquad\qquad\qquad\qquad\qquad\qquad\qquad\qquad
{}\times
\rme^{\rmi(\omega_{k_1}+\cdots+\omega_{k_{m_+}}-\omega_{k_1'}-\cdots-\omega_{k_{m_-}'})t}
.
\end{align}
\end{widetext}
As we discussed below (\ref{density1}), not only the spectrum $K_{\bm{k}_1\ldots\bm{k}_{m_+}\bm{k}_1'\ldots\bm{k}_{m_-}'}$ of the correlation function but also observables are relevant to the evolution of the cumulant through $J_{\bm{k}}$.
To generate the expectation value of an observable, we take the derivatives of the characteristic functional $\chi_t[J,J^*]$ with respect to $J_{\bm{k}}$ and $J_{\bm{k}}^*$, through which $J_{\bm{k}}$ and $J_{\bm{k}}^*$ are replaced by the Fourier spectrum of the observable \cite{note:Locality}.
In this sense, the $\bm{k}$ dependence of $J_{\bm{k}}$ represents the spectra of observables of interest.
Let us consider a specific form of the overall spectrum for the evolution of the cumulant, i.e.\ a Gaussian spectrum with a quadratic dispersion relation $\omega_k=\frac{1}{2}a^2k^2$, to facilitate explicit calculation and to get an idea on how the cumulants decay.
In the case of a symmetric Gaussian spectrum, the local component evolves as
\begin{widetext}
\begin{align}
\tilde{\mathcal{K}}_{m_+,m_-}(t)
&=
\int \rmd^D\bm{k}_1
\cdots
\int \rmd^D\bm{k}_{m_+}
\int \rmd^D\bm{k}_1'
\cdots
\int \rmd^D\bm{k}_{m_-}'
\tilde{K}_{\bm{k}_1\ldots\bm{k}_{m_+}\bm{k}_1'\ldots\bm{k}_{m_-}'}
J_{\bm{k}_1}\cdots J_{\bm{k}_{m_+}}
J_{\bm{k}_1'}^*\cdots J_{\bm{k}_{m_-}'}^*
\nonumber\\[-1truemm]
&\qquad\qquad\qquad\qquad\qquad\qquad\qquad\qquad\qquad\qquad\qquad\qquad\qquad\qquad
{}\times
\rme^{\rmi(\omega_{k_1}+\cdots+\omega_{k_{m_+}}-\omega_{k_1'}-\cdots-\omega_{k_{m_-}'})t}
\nonumber
\displaybreak[0]
\\
&
\propto
\int \rmd^D\bm{k}_1
\cdots
\int \rmd^D\bm{k}_{m_+}
\int \rmd^D\bm{k}_1'
\cdots
\int \rmd^D\bm{k}_{m_-}'
\rme^{-\frac{1}{2}\sigma^2(\bm{k}_1^2+\cdots+\bm{k}_{m_+}^2+\bm{k}_1'^2+\cdots+\bm{k}_{m_-}'^2)}
\rme^{\frac{\rmi}{2}a^2(\bm{k}_1^2+\cdots+\bm{k}_{m_+}^2-\bm{k}_1'^2-\cdots-\bm{k}_{m_-}'^2)t}
\nonumber
\displaybreak[0]
\\
&
=
\frac{(2\pi)^{(m_++m_-)D/2}}{(\sigma^2-\rmi a^2t)^{m_+D/2}(\sigma^2+\rmi a^2t)^{m_-D/2}}.
\end{align}
The translationally invariant component with such a symmetric Gaussian spectrum, on the other hand, evolves as
\begin{align}
\bar{\mathcal{K}}_{m_+,m_-}(t)
&=
\int \rmd^D\bm{k}_1
\cdots
\int \rmd^D\bm{k}_{m_+}
\int \rmd^D\bm{k}_1'
\cdots
\int \rmd^D\bm{k}_{m_-}'
\bar{K}_{\bm{k}_1\ldots\bm{k}_{m_+}\bm{k}_1'\ldots\bm{k}_{m_-}'}
\delta^D(\bm{k}_1+\cdots+\bm{k}_{m_+}-\bm{k}_1'-\cdots-\bm{k}_{m_-}')
\nonumber\\
&\qquad\qquad\qquad\qquad\qquad\qquad\qquad\qquad\qquad\qquad\ \ %
{}\times
J_{\bm{k}_1}\cdots J_{\bm{k}_{m_+}}
J_{\bm{k}_1'}^*\cdots J_{\bm{k}_{m_-}'}^*
\rme^{\rmi(\omega_{k_1}+\cdots+\omega_{k_{m_+}}-\omega_{k_1'}-\cdots-\omega_{k_{m_-}'})t}
\nonumber\displaybreak[0]\\
&
\propto
\int \rmd^D\bm{k}_1
\cdots
\int \rmd^D\bm{k}_{m_+}
\int \rmd^D\bm{k}_1'
\cdots
\int \rmd^D\bm{k}_{m_-}'
\delta^D(\bm{k}_1+\cdots+\bm{k}_{m_+}-\bm{k}_1'-\cdots-\bm{k}_{m_-}')
\nonumber\\
&\qquad\qquad\qquad\qquad\qquad\qquad\qquad\qquad\qquad\ %
{}\times
\rme^{-\frac{1}{2}\sigma^2(\bm{k}_1^2+\cdots+\bm{k}_{m_+}^2+\bm{k}_1'^2+\cdots+\bm{k}_{m_-}'^2)}
\rme^{\frac{\rmi}{2}a^2(\bm{k}_1^2+\cdots+\bm{k}_{m_+}^2-\bm{k}_1'^2-\cdots-\bm{k}_{m_-}'^2)t}
\nonumber\displaybreak[0]\\
&
=
\frac{(2\pi)^{(m_++m_--1)D/2}}{
(\sigma^2-\rmi a^2t)^{(m_+-1)D/2}(\sigma^2+\rmi a^2t)^{(m_--1)D/2}
[
(m_++m_-)
\sigma^2
+\rmi(m_+-m_-)a^2t
]^{D/2}
}.
\end{align}
\end{widetext}
The local component of any $m$th-order cumulant ($m=m_++m_-$) decays asymptotically as 
\begin{equation}
\tilde{\mathcal{K}}_{m_+,m_-}(t)
\sim
t^{-mD/2},
\label{eqn:AsympDecayLocal}
\end{equation}
while the translationally invariant component as 
\begin{equation}
\bar{\mathcal{K}}_{m_+,m_-}(t)
\sim
\begin{cases}
\medskip
t^{-(m-1)D/2}&(m_+\neq m_-),\\
t^{-(m-2)D/2}&(m_+=m_-).
\end{cases}
\label{eqn:AsympDecayNonLocal}
\end{equation}
These show that (i) the cumulants decay except for the translationally invariant component $f_{\bm{k}}$ of the normal correlation $(m_+,m_-)=(1,1)$.
This leads to the Gaussification.
In addition, we see that (ii) the translationally invariant components decay more slowly than the local components, and that (iii) the decays of the translationally invariant components of the gauge-invariant cumulants with $m_+=m_-$ are even slower.

Even if the real spectra for the evolutions of the cumulants are not of such simple Gaussian form, the saddle-point approximation in estimating the integrals for large $t$ yields Gaussian integrals.
The integrands after the saddle-point approximation might not be so simple as the symmetric Gaussian considered above, but in any case, for generic (not necessarily symmetric) Gaussian spectra with the quadratic dispersion relation $\omega_k$, the cumulants decay asymptotically as (\ref{eqn:AsympDecayLocal}) and (\ref{eqn:AsympDecayNonLocal}).
The results (\ref{eqn:AsympDecayLocal}) and (\ref{eqn:AsympDecayNonLocal}) are therefore of quite general validity.

\section{Mixing of GGE}
\label{app:Mixing}
Here, we prove the mixing property (\ref{mixing1}) of the GGE in (\ref{density1}) with the Hamiltonian $\hat{H}$ in (\ref{eqn:FreeBoson}).
Recall the characteristic functional of the GGE in (\ref{Characteristic2}), \begin{align}
	\chi_\text{GGE}[J,J^*]
	&=\langle\hat{W}[J,J^*]\rangle_\text{GGE}
	\nonumber\displaybreak[0]\\
	&=\exp\!\left(
	-\frac{1}{2}\int\rmd^D\bm{k}\,(1+2f_{\bm{k}})|J_{\bm{k}}|^2
	\right),
	\label{eqn:CharGGE}
\end{align}
where
\begin{equation}
\hat{W}[J,J^*]
=\rme^{\int\rmd^D\bm{k}\,(J_{\bm{k}}\hat{b}_{\bm{k}}^\dag-J_{\bm{k}}^*\hat{b}_{\bm{k}})},
\end{equation}
and consider
\begin{align}
&\Xi_t[J^{(A)},J^{(A)*},J^{(B)},J^{(B)*},J^{(C)},J^{(C)*}]
\nonumber\displaybreak[0]\\
&\quad
=
\Bigl\langle
\hat{W}[J^{(A)},J^{(A)*}]
\rme^{\rmi\hat{H}t}
\hat{W}[J^{(B)},J^{(B)*}]
\rme^{-\rmi\hat{H}t}
\nonumber\displaybreak[0]\\[-1.5truemm]
&\qquad\qquad\qquad\qquad\qquad\qquad\quad
{}\times
\hat{W}[J^{(C)},J^{(C)*}]
\Bigr\rangle_\text{GGE},
\end{align}
which is the generating functional for the correlation functions of the type (\ref{mixing1}).
In the GGE characterized by (\ref{eqn:CharGGE}), it reads
\begin{widetext}
\begin{align}
&\Xi_t[J^{(A)},J^{(A)*},J^{(B)},J^{(B)*},J^{(C)},J^{(C)*}]
\nonumber\displaybreak[0]\\
&\qquad
=
\Bigl\langle
\hat{W}[J^{(A)},J^{(A)*}]
\hat{W}[J^{(B)}\rme^{\rmi\omega t},J^{(B)*}\rme^{-\rmi\omega t}]
\hat{W}[J^{(C)},J^{(C)*}]
\Bigr\rangle_\text{GGE}
\nonumber\displaybreak[0]\\
&\qquad
=\exp\!\left(
-\rmi\Im\int\rmd^D\bm{k}\,(J_{\bm{k}}^{(A)*}J_{\bm{k}}^{(B)}\rme^{\rmi\omega_kt}+J_{\bm{k}}^{(B)*}J_{\bm{k}}^{(C)}\rme^{-\rmi\omega_kt}+J_{\bm{k}}^{(A)*}J_{\bm{k}}^{(C)})
\right)
\nonumber\displaybreak[0]\\
&\qquad\qquad\qquad\qquad\qquad\qquad
{}\times
\left\langle
\hat{W}[J^{(A)}+J^{(B)}\rme^{\rmi\omega t}+J^{(C)},J^{(A)*}+J^{(B)*}\rme^{-\rmi\omega t}+J^{(C)*}]
\right\rangle_\text{GGE}
\nonumber\displaybreak[0]\\
&\qquad
=\exp\!\left(
-\rmi\Im\int\rmd^D\bm{k}\,(J_{\bm{k}}^{(A)*}J_{\bm{k}}^{(B)}\rme^{\rmi\omega_kt}+J_{\bm{k}}^{(B)*}J_{\bm{k}}^{(C)}\rme^{-\rmi\omega_kt}+J_{\bm{k}}^{(A)*}J_{\bm{k}}^{(C)})
\right)
\nonumber\displaybreak[0]\\
&\qquad\qquad\qquad\qquad\qquad\qquad
{}\times
\exp\!\left(
-\frac{1}{2}\int\rmd^D\bm{k}\,(1+2f_{\bm{k}})|J_{\bm{k}}^{(A)}+J_{\bm{k}}^{(B)}\rme^{\rmi\omega_kt}+J_{\bm{k}}^{(C)}|^2
\right).
\intertext{This relaxes to}
&\Xi_t[J^{(A)},J^{(A)*},J^{(B)},J^{(B)*},J^{(C)},J^{(C)*}]
\nonumber\displaybreak[0]\\
&\qquad
\xrightarrow{t\to\infty}
\exp\!\left(
-\rmi\Im\int\rmd^D\bm{k}\,J_{\bm{k}}^{(A)*}J_{\bm{k}}^{(C)}
\right)
\exp\!\left(
-\frac{1}{2}\int\rmd^D\bm{k}\,(1+2f_{\bm{k}})(|J_{\bm{k}}^{(A)}+J_{\bm{k}}^{(C)}|^2+|J_{\bm{k}}^{(B)}|^2)
\right)
\nonumber\displaybreak[0]\\
&\qquad
=
\left\langle
\hat{W}[J^{(A)},J^{(A)*}]
\hat{W}[J^{(C)},J^{(C)*}]
\right\rangle_\text{GGE}
\left\langle
\hat{W}[J^{(B)},J^{(B)*}]
\right\rangle_\text{GGE},
\end{align}
\end{widetext}
by the Riemann-Lebesgue lemma.
This shows the factorization (\ref{mixing1}) for any local observables $\hat{A}$, $\hat{B}$, and $\hat{C}$.

\section{Relaxation from Correlated Initial States}
\label{app:RelaxCorrelated}
Here, we prove the relaxation (\ref{eqn:RelaxCorrelated}) from the correlated initial state (\ref{eqn:CorrelatedInitialState}).
We actually look at the evolution of the characteristic functional of the state of the total system driven by the Hamiltonian $\hat{H}$ in (\ref{diagonal1}),
\begin{equation}
\chi_t[J,J^*]
=\langle\hat{W}[J,J^*]\rangle_t
=\sum_j\langle\hat{L}_j^\dag\rme^{\rmi\hat{H}t}\hat{W}[J,J^*]\rme^{-\rmi\hat{H}t}\hat{L}_j\rangle,
\label{eqn:CharFuncCorrelatedInitStat}
\end{equation}
where
\begin{equation}
\hat{W}[J,J^*]
=\rme^{\sum_\ell\int\rmd^D\bm{k}\,(J_{\bm{k}\ell}\hat{A}_{\bm{k}\ell}^\dag-J_{\bm{k}\ell}^*\hat{A}_{\bm{k}\ell})},
\end{equation}
and $\langle{}\cdots{}\rangle$ denotes the expectation value in the factorized state $\hat{\rho}_S\otimes\hat{\rho}_B$ in (\ref{eqn:InitialProductState}).
We introduce the generating functional for the correlation functions in (\ref{eqn:CharFuncCorrelatedInitStat}),
\begin{align}
&\Xi_t[J,J^*,J^{(L)},J^{(L)*},J^{(L^\dag)},J^{(L^\dag)*}]
\nonumber\displaybreak[0]\\
&\ %
=
\left\langle
\hat{W}[J^{(L)},J^{(L)*}]
\rme^{\rmi\hat{H}t}
\hat{W}[J,J^*]
\rme^{-\rmi\hat{H}t}
\hat{W}[J^{(L^\dag)},J^{(L^\dag)*}]
\right\rangle.
\end{align}
It is reduced to
\begin{widetext}
\begin{align}
&\Xi_t[J,J^*,J^{(L)},J^{(L)*},J^{(L^\dag)},J^{(L^\dag)*}]
\nonumber
%\displaybreak[0]
\\
&\qquad
=
\left\langle
\hat{W}[J^{(L)},J^{(L)*}]
\hat{W}[J\rme^{\rmi\omega t},J^*\rme^{-\rmi\omega t}]
\hat{W}[J^{(L^\dag)},J^{(L^\dag)*}]
\right\rangle
\nonumber
\displaybreak[0]
\\
&\qquad
=\exp\!\left(
-\rmi\Im\sum_\ell\int\rmd^D\bm{k}\,(J_{\bm{k}\ell}^{(L)*}J_{\bm{k}\ell}\rme^{\rmi\omega_{k\ell}t}+J_{\bm{k}\ell}^*J_{\bm{k}\ell}^{(L^\dag)}\rme^{-\rmi\omega_{k\ell}t}+J_{\bm{k}\ell}^{(L)*}J_{\bm{k}\ell}^{(L^\dag)})
\right)
\nonumber
\displaybreak[0]
\\
&\qquad\qquad\qquad\qquad\qquad\qquad\qquad\qquad\qquad\qquad\qquad
{}\times
\left\langle
\hat{W}[J^{(L)}+J\rme^{\rmi\omega t}+J^{(L^\dag)},J^{(L)*}+J^*\rme^{-\rmi\omega t}+J^{(L^\dag)*}]
\right\rangle
\nonumber\displaybreak[0]\\
&\qquad
=\exp\!\left(
-\rmi\Im\sum_\ell\int\rmd^D\bm{k}\,(J_{\bm{k}\ell}^{(L)*}J_{\bm{k}\ell}\rme^{\rmi\omega_{k\ell}t}+J_{\bm{k}\ell}^*J_{\bm{k}\ell}^{(L^\dag)}\rme^{-\rmi\omega_{k\ell}t}+J_{\bm{k}\ell}^{(L)*}J_{\bm{k}\ell}^{(L^\dag)})
\right)
\chi_S\bm{(}\xi(t),\xi^*(t)\bm{)}
\chi_B[\eta(t),\eta^*(t)],
\end{align}
\end{widetext}
where $\chi_S(\xi,\xi^*)$ and $\chi_B[\eta,\eta^*]$ are the characteristic function of the state $\hat{\rho}_S$ of the harmonic oscillator in (\ref{eqn:CharFuncS}) and the characteristic functional of the state $\hat{\rho}_B$ of the reservoirs in (\ref{eqn:CharFuncB}), respectively, with
\begin{gather}
\xi(t)
=\sum_\ell\int\rmd^D\bm{k}\,\alpha_{\bm{k}\ell}^*
(J_{\bm{k}\ell}^{(L)}+J_{\bm{k}\ell}\rme^{\rmi\omega_{k\ell}t}+J_{\bm{k}\ell}^{(L^\dag)}),
\displaybreak[0]\\
\eta_{\bm{k}\ell}(t)
=\sum_{\ell'}\int \rmd^D\bm{k}'\,
\beta_{\bm{k}'\ell',\bm{k}\ell}^*
(J_{\bm{k}'\ell'}^{(L)}+J_{\bm{k}'\ell'}\rme^{\rmi\omega_{k'\ell'}t}+J_{\bm{k}'\ell'}^{(L^\dag)}).
\end{gather}
For $t\to\infty$, due to the Riemann-Lebesgue lemma, it behaves asymptotically as
\begin{align}
&\Xi_t[J,J^*,J^{(L)},J^{(L)*},J^{(L^\dag)},J^{(L^\dag)*}]
\nonumber
\displaybreak[0]
\\
&\ \ %
\xrightarrow{t\to\infty}
\exp\!\left(
-\rmi\Im\sum_\ell\int\rmd^D\bm{k}\,J_{\bm{k}\ell}^{(L)*}J_{\bm{k}\ell}^{(L^\dag)}
\right)
\nonumber
%\displaybreak[0]
\\
&\qquad\qquad\qquad\qquad\qquad
{}\times
\chi_S(\bar{\xi},\bar{\xi}^*)
\chi_B[\eta(t),\eta^*(t)],
\label{eqn:ChiCorrelatedAsymp}
\end{align}
with
\begin{align}
\xi(t)
&
\to
\sum_\ell\int\rmd^D\bm{k}\,\alpha_{\bm{k}\ell}^*
(J_{\bm{k}\ell}^{(L)}+J_{\bm{k}\ell}^{(L^\dag)})
=\bar{\xi},
\displaybreak[0]\\
\eta_{\bm{k}\ell}(t)
&
\to
\sum_{\ell'}\int \rmd^D\bm{k}'\,
\beta_{\bm{k}'\ell',\bm{k}\ell}^*
(J_{\bm{k}'\ell'}^{(L)}+J_{\bm{k}'\ell'}^{(L^\dag)})
+J_{\bm{k}\ell}\rme^{\rmi\omega_{k\ell}t}
\nonumber\displaybreak[0]\\
&=
\bar{\eta}_{\bm{k}\ell}
+J_{\bm{k}\ell}\rme^{\rmi\omega_{k\ell}t}
.
\end{align}
Recall (\ref{eqn:XiDecay}) and (\ref{eqn:EtaDecay}) for the factorized case.
Inserting the asymptotic behavior of $\eta_{\bm{k}\ell}(t)$ in the cumulant expansion of $\chi_B[\eta,\eta^*]$ in (\ref{eqn:CumExpB}), we have
\begin{widetext}
\begin{align}
&\chi_B[\eta(t),\eta^*(t)]
\nonumber\\
&\ %
=
\exp\biggl(
2\rmi\Im\sum_\ell
\int \rmd^D\bm{k}\,
(
\bar{\eta}_{\bm{k}\ell}
+J_{\bm{k}\ell}\rme^{\rmi\omega_{k\ell}t}
)
\langle\hat{b}_{\bm{k}}^\dag\rangle
\nonumber\displaybreak[0]\\
&\qquad\quad\ \ %
{}
-\frac{1}{2}
\sum_\ell
\int \rmd^D\bm{k}\,
(1+2f_{\bm{k}}^{(\ell)})|
\bar{\eta}_{\bm{k}\ell}
+J_{\bm{k}\ell}\rme^{\rmi\omega_{k\ell}t}
|^2
\nonumber\displaybreak[0]\\
&\qquad\quad\ \ %
{}
+
\Re\sum_\ell\int \rmd^D\bm{k}\,
g_{\bm{k}}^{(\ell)*}
(
\bar{\eta}_{\bm{k}\ell}
+J_{\bm{k}\ell}\rme^{\rmi\omega_{k\ell}t}
)(
\bar{\eta}_{-\bm{k}\ell}
+J_{-\bm{k}\ell}\rme^{\rmi\omega_{k\ell}t}
)
\nonumber\displaybreak[0]\\
&\qquad\quad\ \ %
{}
-
\sum_\ell
\int \rmd^D\bm{k}
\int \rmd^D\bm{k}'\,
(
\bar{\eta}_{\bm{k}\ell}
+J_{\bm{k}\ell}\rme^{\rmi\omega_{k\ell}t}
)
F_{\bm{k}\bm{k}'}^{(\ell)}
(
\bar{\eta}_{\bm{k}'\ell}^*
	+J_{\bm{k}'\ell}^*\rme^{-\rmi\omega_{k'\ell}t}
)
\nonumber\displaybreak[0]\\
&\qquad\quad\ \ %
{}+
\Re
\sum_\ell
\int \rmd^D\bm{k}
\int \rmd^D\bm{k}'\,
(
\bar{\eta}_{\bm{k}\ell}
+J_{\bm{k}\ell}\rme^{\rmi\omega_{k\ell}t}
)
G_{\bm{k}\bm{k}'}^{(\ell)*}
(
\bar{\eta}_{\bm{k}'\ell}
+J_{\bm{k}'\ell}\rme^{\rmi\omega_{k'\ell}t}
)
\nonumber\displaybreak[0]\\
&\qquad\quad\ \ %
{}+\sum_\ell
\int \rmd^D\bm{k}_1
\int \rmd^D\bm{k}_2
\int \rmd^D\bm{k}_3\,
K_{\bm{k}_1\bm{k}_2\bm{k}_3}^{(\ell)}
(
\bar{\eta}_{\bm{k}_1\ell}
+J_{\bm{k}_1\ell}\rme^{\rmi\omega_{k_1\ell}t}
)
(
\bar{\eta}_{\bm{k}_2\ell}
+J_{\bm{k}_2\ell}\rme^{\rmi\omega_{k_2\ell}t}
)
(
\bar{\eta}_{\bm{k}_3\ell}^*
+J_{\bm{k}_3\ell}^*\rme^{-\rmi\omega_{k_3\ell}t}
)
+\cdots
\biggr)
\nonumber\displaybreak[0]\\
&\ %
\xrightarrow{t\to\infty}
\exp\biggl(
2\rmi\Im\sum_\ell
\int \rmd^D\bm{k}\,
\bar{\eta}_{\bm{k}\ell}
\langle\hat{b}_{\bm{k}}^\dag\rangle
\nonumber\displaybreak[0]\\
&\qquad\qquad\quad\ \ %
{}-\frac{1}{2}
\sum_\ell
\int \rmd^D\bm{k}\,
(1+2f_{\bm{k}}^{(\ell)})
(
|
\bar{\eta}_{\bm{k}\ell}
|^2
+|J_{\bm{k}\ell}|^2
)
+
\Re\sum_\ell\int \rmd^D\bm{k}\,
g_{\bm{k}}^{(\ell)*}
\bar{\eta}_{\bm{k}\ell}
\bar{\eta}_{-\bm{k}\ell}
\nonumber\displaybreak[0]\\
&\qquad\qquad\quad\ \ %
{}-
\sum_\ell
\int \rmd^D\bm{k}
\int \rmd^D\bm{k}'\,
\bar{\eta}_{\bm{k}\ell}
F_{\bm{k}\bm{k}'}^{(\ell)}
\bar{\eta}_{\bm{k}'\ell}^*
+
\Re
\sum_\ell
\int \rmd^D\bm{k}
\int \rmd^D\bm{k}'\,
\bar{\eta}_{\bm{k}\ell}
G_{\bm{k}\bm{k}'}^{(\ell)*}
\bar{\eta}_{\bm{k}'\ell}
\nonumber\displaybreak[0]\\
&\qquad\qquad\quad\ \ %
{}+\sum_\ell
\int \rmd^D\bm{k}_1
\int \rmd^D\bm{k}_2
\int \rmd^D\bm{k}_3\,
K_{\bm{k}_1\bm{k}_2\bm{k}_3}^{(\ell)}
\bar{\eta}_{\bm{k}_1\ell}
\bar{\eta}_{\bm{k}_2\ell}
\bar{\eta}_{\bm{k}_3\ell}^*
+\cdots
\biggr)
\nonumber\displaybreak[0]\\
&\ %
=\chi_B[
\bar{\eta}
,
\bar{\eta}^*
]
\chi_\text{NESS}[J,J^*],
\end{align}
\end{widetext}
where $\chi_\text{NESS}[J,J^*]$ is given in (\ref{eqn:GNESS}).
Thus, (\ref{eqn:ChiCorrelatedAsymp}) further relaxes to
\begin{align}
&\Xi_t[J,J^*,J^{(L)},J^{(L)*},J^{(L^\dag)},J^{(L^\dag)*}]
\nonumber
\displaybreak[0]
\\
&\ \ %
\xrightarrow{t\to\infty}
\exp\!\left(
-\rmi\Im\sum_\ell\int\rmd^D\bm{k}\,J_{\bm{k}\ell}^{(L)*}J_{\bm{k}\ell}^{(L^\dag)}
\right)
\nonumber
%\displaybreak[0]
\\
&\qquad\qquad\qquad
{}\times
\chi_S(\bar{\xi},\bar{\xi}^*)
\chi_B[
\bar{\eta}
,
\bar{\eta}^*
]
\chi_\text{NESS}[J,J^*]
\nonumber
\displaybreak[0]
\\
&\ \ %
=\left\langle
\hat{W}[J^{(L)},J^{(L)*}]
\hat{W}[J^{(L^\dag)},J^{(L^\dag)*}]
\right\rangle
\chi_\text{NESS}[J,J^*].
\end{align}
This shows that the characteristic functional (\ref{eqn:CharFuncCorrelatedInitStat}) for the correlated initial state $\hat{\rho}_0$ in (\ref{eqn:CorrelatedInitialState}) relaxes to
\begin{equation}
\chi_t[J,J^*]
\xrightarrow{t\to\infty}
\sum_j\langle\hat{L}_j^\dag\hat{L}_j\rangle
\chi_\text{NESS}[J,J^*]
\end{equation}
for any local perturbations $\hat{L}_j$, and proves (\ref{eqn:RelaxCorrelated}), under the normalization condition $\sum_j\langle\hat{L}_j^\dag\hat{L}_j\rangle=1$ for the correlated initial state $\hat{\rho}_0$.

%\bibliography{FriedrichsGGE,kazuya}

\begin{thebibliography}{10}
\bibitem{note:QuantumIntegrability}
The notion of integrability for quantum systems is discussed in Refs.\
  \cite{ref:Caux-QuantumIntegrability,ref:Braak-RabiExact}, but we do not go
  into this issue in the present work.
\bibitem{note:Matrix}
For instance, the product $VV'$ of $V_{\bm{k}\bm{k}'}$ and $V'_{\bm{k}\bm{k}'}$
  is defined by
  $(VV')_{\bm{k}\bm{k}'}={\protect\int{}}\rmd^D\bm{k}''\,V_{\bm{k}\bm{k}''}V'_{\bm{k}''\bm{k}'}$
  and the inverse $V^{-1}$ by
  ${\protect\int{}}\rmd^D\bm{k}''\,(V^{-1})_{\bm{k}\bm{k}''}V_{\bm{k}''\bm{k}'}={\protect\int{}}\rmd^D\bm{k}''\,V_{\bm{k}\bm{k}''}(V^{-1})_{\bm{k}''\bm{k}'}=\delta^D(\bm{k}-\bm{k}')\openone_2$,
  where $\openone_2$ is the $2\times2$ identity matrix.
\bibitem{note:Locality}
Consider for instance a quadratic operator
  $\hat{A}={\protect\int{}}\rmd^D\bm{k}{{}\protect\int{}}\rmd^D\bm{k}'\,\mathcal{A}_{\bm{k}\bm{k}'}\hat{b}_{\bm{k}}^\dag\hat{b}_{\bm{k}'}$.
  If the Fourier coefficient $\mathcal{A}_{\bm{k}\bm{k}'}$ contains
  $\delta^D(\bm{k}-\bm{k}')$, it represents a translationally invariant
  component in space, and this quantity spreads all over the space. We do not
  consider such nonlocal quantities. Local observables are free from such delta
  singularity, which inhibits the decay of Fourier transforms. This condition
  is inherited by the regularity (free from delta singularity) of $J_{\bm{k}}$
  in the characteristic functionals. We are not interested in nonlocal
  observables, since the expectation values of such quantities diverge, and
  they are not ``observable'' experimentally.
\bibitem{note:FormFactor}
The form factor $u_{\bm{k}\ell}$ characterizes the spatial region where the harmonic oscillator interacts with the $\ell$th reservoir. Since the wave functions of the harmonic oscillator have some finite widths, the interaction is not point-like. This width cannot be neglected. If we neglect it, assuming a point-like interaction, we suffer from unphysical divergence in the field-theoretical calculations.
\end{thebibliography}
%\end{document}

\end{document}